\documentclass[11pt,twoside]{article}
\usepackage{asp2010}

\resetcounters

\aspvolume{???}
\aspcpryear{2010}
\aspvoltitle{The Dynamic Interstellar Medium: A Celebration of the Canadian Galactic Plane Survey}
\aspvolauthor{R. Kothes and T. L. Landecker}

\begin{document}

%\marginpar{\bf 
%\footnotesize
%Full-resolution grayscale \\
%and color versions of this \\
%preprint are available at \\
%physics.wku.edu/$^\sim$gibson 
%}

\title{Cold Atomic Gas in the CGPS and Beyond}
\author{Steven J. Gibson$^1$}
\affil{$^1$Dept.\ of Physics \& Astronomy, Western Kentucky University, 1906 College Heights Blvd., Bowling Green, KY 42101, USA}

\begin{abstract}
The Canadian Galactic Plane Survey has opened new vistas on the Milky Way,
including cold hydrogen clouds that bridge a critical gap between the
classical diffuse interstellar medium and the gravitationally bound molecular
clouds that can form stars.  The CGPS and its fellow IGPS surveys revealed
these transitional clouds to be surprisingly widespread as {\sc H~i}
self-absorption (HISA) shadows against the Galactic {\sc H~i} emission
background.  The richness of the IGPS data allows detailed examination of
HISA cloud spatial structure, gas properties, Galactic distribution, and
correspondence with molecular gas, all of which can constrain models of cold
{\sc H~i} clouds in the evolving interstellar medium.  Augmenting the
landmark IGPS effort are new and upcoming surveys with the Arecibo 305m and
Australian SKA Pathfinder telescopes.
\end{abstract}

\section{Observational Context}
\label{Sec:obs}

This review is of limited scope in order to leave room for a few current
results.  
Readers are encouraged to consult an earlier
review \citep{gibson_2002}, as well as excellent broader {\sc H~i} reviews by
\citet{kh_1988}, \citet{dl_1990}, and \citet{kalberla_2009}, still-broader
ISM reviews by \citet{wolfire_1995,wolfire_2003}, \citet{cox_2005},
and \citet{snow_2006}, 
and many other articles in these proceedings.

Neutral atomic hydrogen ({\sc H~i}), the dominant constituent of interstellar
matter in the Galactic disk, is found in a broad range of environments, from
diffuse gas with $T \sim 10^3 - 10^4$~K (the warm neutral medium = WNM) to
cold clouds with $T \sim 10 - 10^2$~K (the cold neutral medium = CNM).
Consequently, the {\sc H~i} 21cm line is used to study the structure,
properties, and distribution of gas in both the ambient ISM and denser,
quiescent pockets where H$_2$ forms, the first step toward star formation.
CNM observations, the subject of this review, allow close scrutiny of (1) the
atomic-to-molecular phase transition, (2) intricate cloud structure shaped by
shocks, turbulence, and perhaps magnetic fields, and (3) spiral density waves
that affect {\sc H~i} radiative transfer.  The Canadian, VLA, and Southern
Galactic Plane Surveys (CGPS: \citealt{cgps}; VGPS: \citealt{vgps}; SGPS:
\citealt{sgps}; together, ``the IGPS'') have transformed our view of the CNM
on all of these fronts, opening the way for future work with the next
generation of {\sc H~i} surveys.

%%%%%%%%%%%%%%%%%%%%%%%%%%%%%%%%%%%%%%%%%%%%%%%%%%%%%%%%%%%%%%%%%%%%%%%%%%%%%%
\begin{figure}[htb]
\centerline{
%\plottwo{gibson_steven_f1a_gray.eps}{gibson_steven_f1b_gray.eps}
%\plottwo{gibson_steven_f1a_color.eps}{gibson_steven_f1b_color.eps}
\plottwo{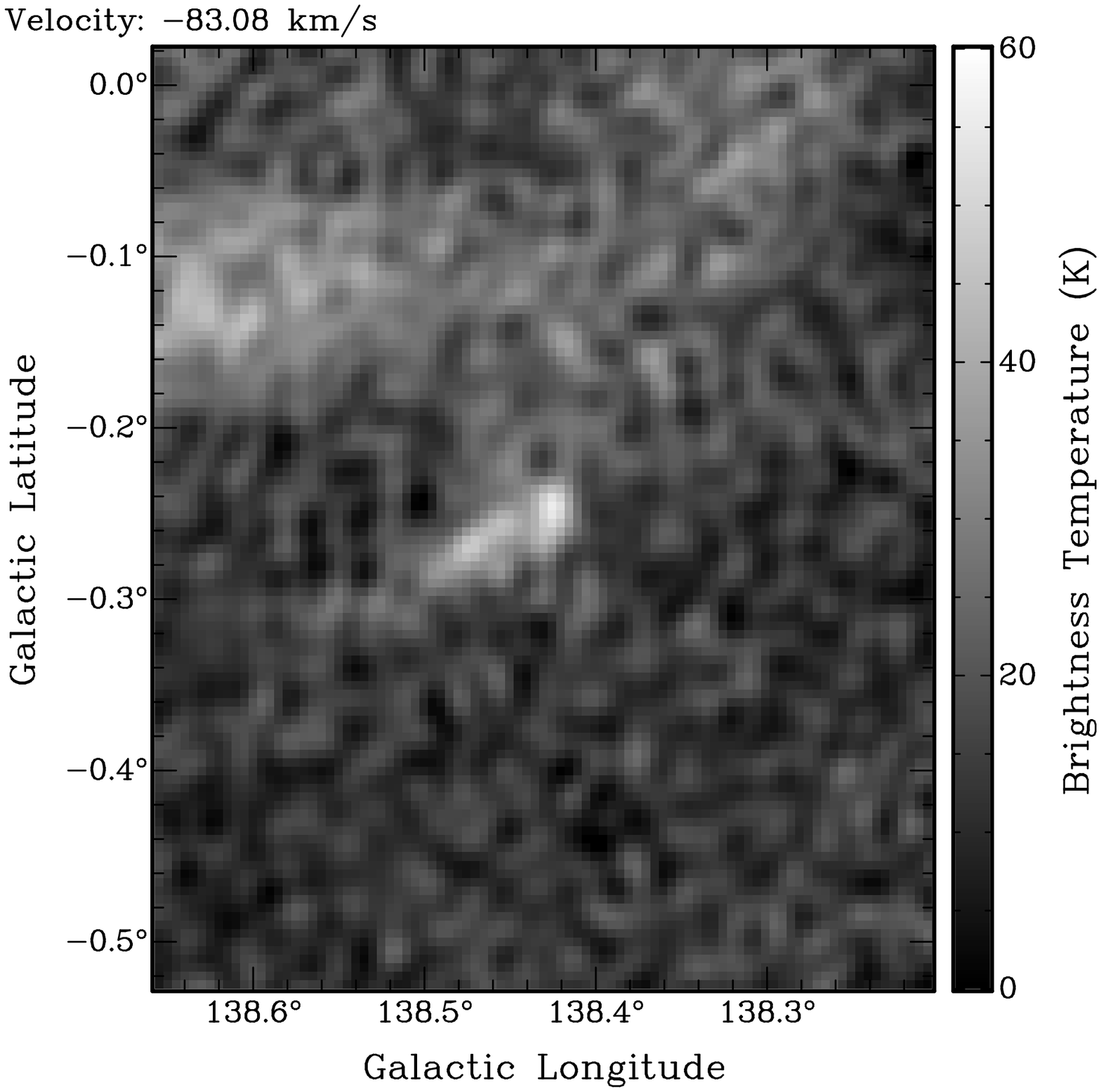}{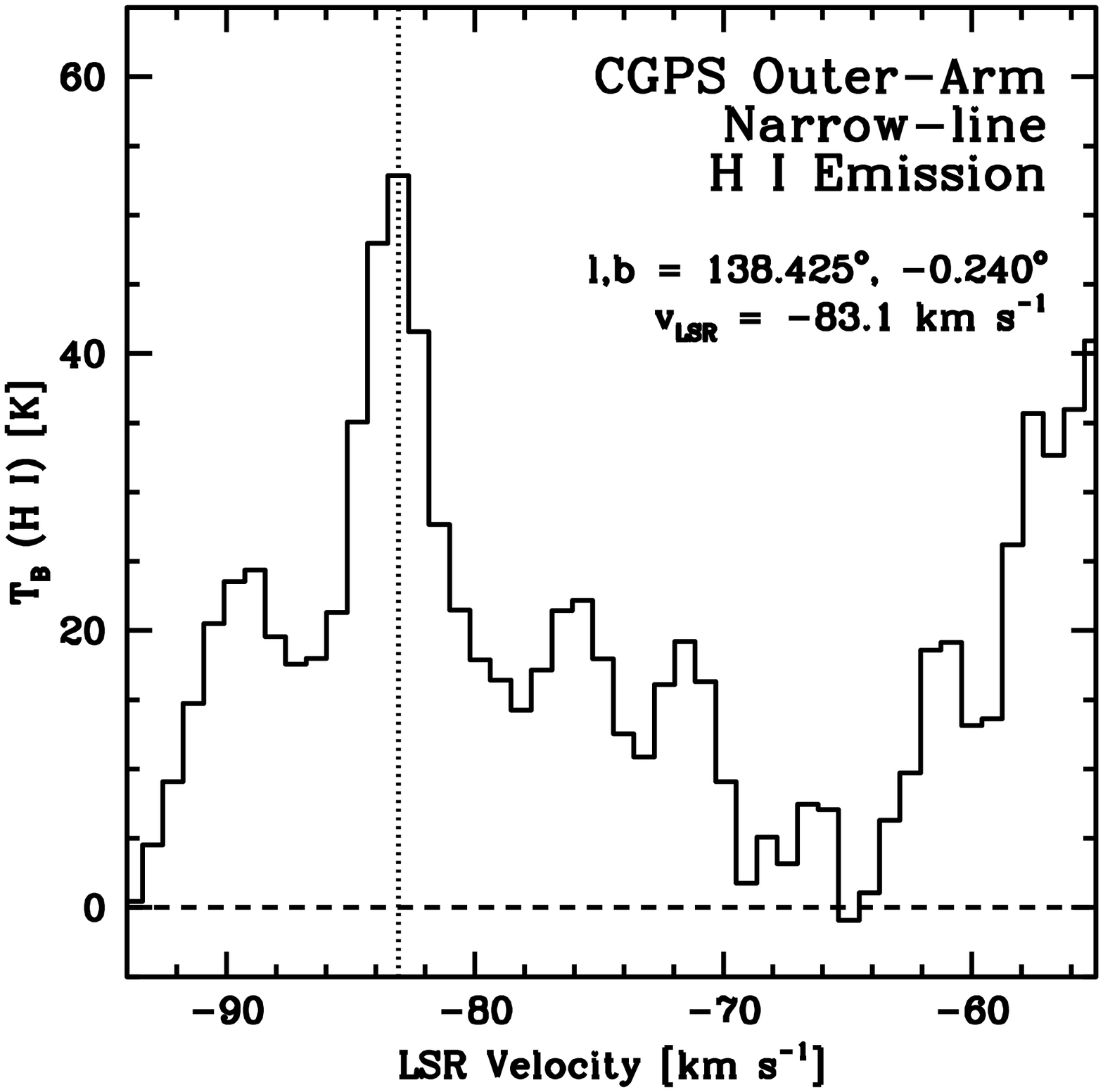}
}
\caption{
CGPS narrow-line {\sc H~i} emission (NHIE) feature in the outer Galaxy 
($d \ga 6$~kpc), 
with
ON-OFF brightness $\Delta T_b = 38$~K,
line width $\Delta v = 2.8$~km/s,
and angular width $\Delta \theta = 2\arcmin \ga 3$~pc.
Gas properties of 
$T_s \sim 80 - 120$~K,
$\tau \sim 0.5 - 1.0$,
$N_{HI} \sim 3 - 5 \times 10^{20}$~cm$^{-2}$,
and
$M_{HI} \ga 25$~M$_\odot$
are consistent with measurements.  No corresponding CO or FIR emission is 
apparent.
}
\label{Fig:cgps_nhie}
\end{figure}
%%%%%%%%%%%%%%%%%%%%%%%%%%%%%%%%%%%%%%%%%%%%%%%%%%%%%%%%%%%%%%%%%%%%%%%%%%%%%%%

A simple demonstration that both WNM and CNM temperature regimes exist is to
compare {\sc H~i} absorption toward a compact continuum source with {\sc H~i}
emission adjacent to the source.  Early interferometric studies of this sort
\citep{clark_1965,radhakrishnan_1972} showed that narrow-line emission
features have matching narrow-line absorption, but broad-line emission lacks
obvious absorption counterparts, except at a very weak level (e.g.,
\citealt{dwarakanath_2002}.)  Two complementary physical arguments apply.
One is that ``warm'' means ''poorly absorbing'': the line center optical
depth varies inversely with temperature as $\tau_0 = C \, N \, / \, (T_s \,
\Delta v)$, where $N$ is column density, $T_s$ is ``spin'' (excitation)
temperature, $\Delta v$ is the line full width at half maximum (FWHM), and $C
= 5.2 \times 10^{-19} \rm cm^2 \, K \, km/s$ for Gaussian lines
\citep{dl_1990}.
The other argument is that ``cold'' means ``narrow line'', or less thermal
broadening: $\Delta v_{therm} = 0.215 \sqrt{T_k}$~km/s for {\sc H~i}, where
$T_k$ is the gas kinetic temperature \citep{spitzer_1978}.  Generally $T_s
\simeq T_k$ in the CNM, but $T_s < T_k$ may occur in the WNM under some
conditions \citep{liszt_2001}.
Although turbulence can broaden CNM lines significantly, it is rarely enough to
confuse them with WNM lines, which form an observationally distinct population
\citep{ht_2003}.

Whether an {\sc H~i} cloud is seen in emission or absorption depends on the
sight line geometry.  For illustration, consider the
simple two-component {\sc H~i} radiative transfer equation (derived in
\citealt{dickey_2002}).  Its observed brightness temperature is

\begin{equation}
\label{Eqn:two_comp}
T_B\,(v) \;\; = \;\; T_s \, \left[ 1 - e^{-\tau(v)} \right] 
\;+\; T_{bg}(v) \; e^{-\tau(v)} 
\;\; ,
\end{equation}

\noindent where 
$v$ is radial velocity, 
$T_s$ and $\tau(v)$ apply to the foreground cloud,
and 
$T_{bg}(v)$ is the background brightness temperature.  
The cloud produces net emission if it is warm relative to the background
brightness ($T_s > T_{bg}$) and absorption if cold ($T_s < T_{bg}$).
The same cloud can also change from emission to absorption against a varying
background.  
Since $\tau \propto {T_s}^{-1}$, WNM and CNM emission features can have similar
brightness despite radically different $T_s$, so these phases are best
distinguished by emission line width.  Sufficiently narrow-line {\sc H~i}
emission (NHIE) traces CNM unambiguously.  For example, $\Delta v \la 4$~km/s
$\Rightarrow T_k \la 300$~K, with the actual $T_k$ probably well below this if
turbulence is present.  Historically only a few NHIE features were known
\citep{kv_1972,goerigk_1983}, but increasingly sophisticated spectral
decompositions have revealed more \citep{vs_1989,poppel_1994,haud_2010}.  In
the CGPS, NHIE features can be found rather easily by eye
(Fig.~\ref{Fig:cgps_nhie}).

Cold {\sc H~i} is more commonly identified in absorption against either a
continuum source ({\sc H~i} continuum absorption = HICA) or other line
emission ({\sc H~i} self-absorption = HISA).  These two approaches are highly
complementary.  HICA is the method of choice for exploring CNM properties
(e.g., \citealt{dickey_2003,ht_2003}), because it has fewer radiative
transfer unknowns.  One can move off the {\sc H~i} frequency to get $T_{bg}$
exactly, where HISA (and NHIE) line backgrounds must be estimated.  One can
also move off the continuum position to see the cloud in emission, a HISA
rarity \citep{kerton_2005}.  But HICA sight lines are discrete and well
separated in present surveys (0.6 per deg$^2$ in the CGPS;
\citealt{st_2004}), so individual clouds are often sampled only once, or
missed entirely.  On larger scales though, HICA sampling in the IGPS is
sufficient to see Galactic structure: \citet{strasser_2007} give a beautiful
longitude-velocity HICA map of spiral arms in CNM, while \citet{dickey_2009}
find a surprisingly steady mix of CNM and WNM in the outer disk, where
equilibrium models predict less CNM or even none due to a drop in pressure
\citep{wolfire_2003}.

\begin{figure}[htb]
\centerline{
%\plottwo{gibson_steven_f2a_gray.eps}{gibson_steven_f2b_gray.eps}
%\plottwo{gibson_steven_f2a_color.eps}{gibson_steven_f2b_color.eps}
\plottwo{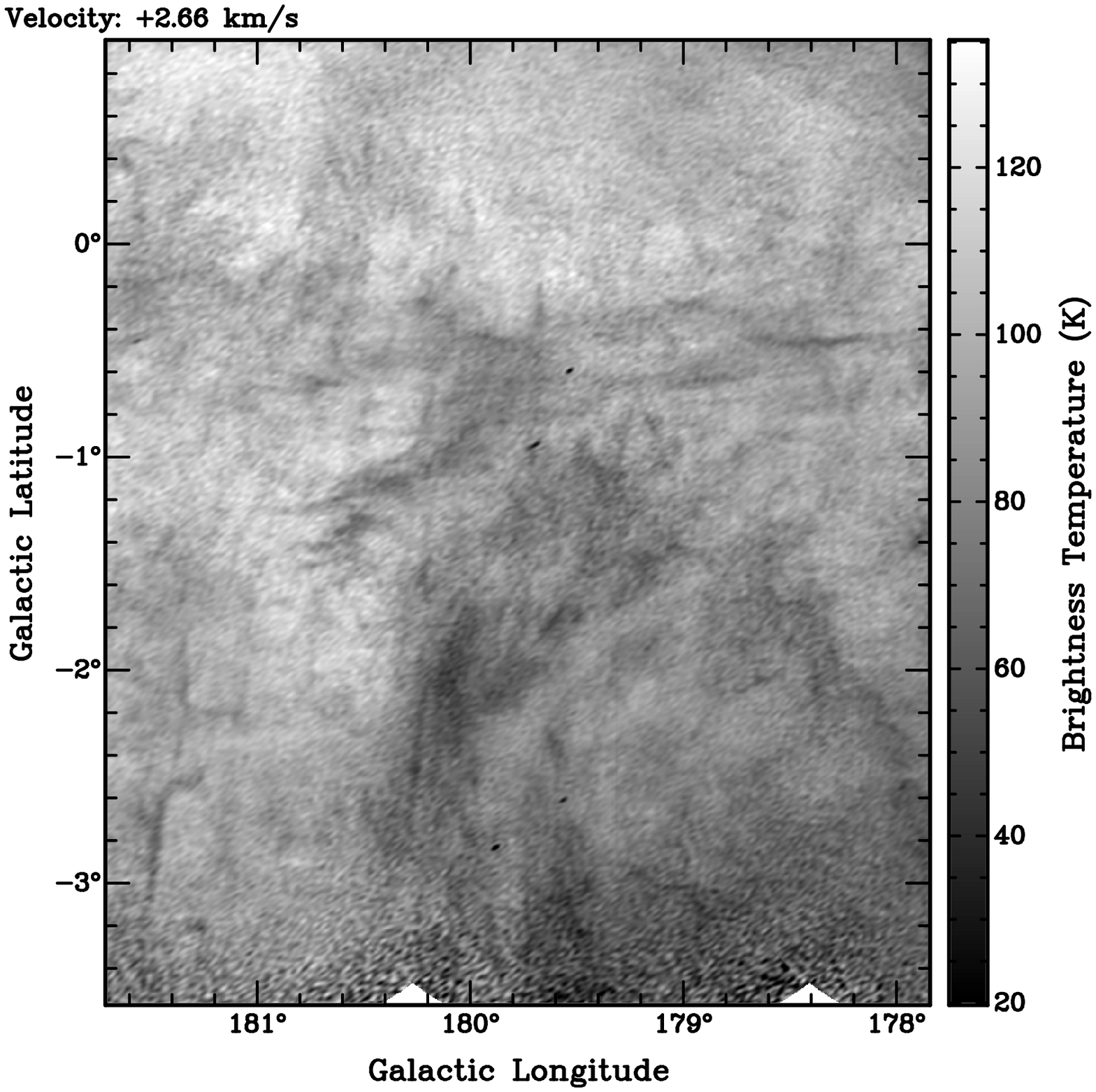}{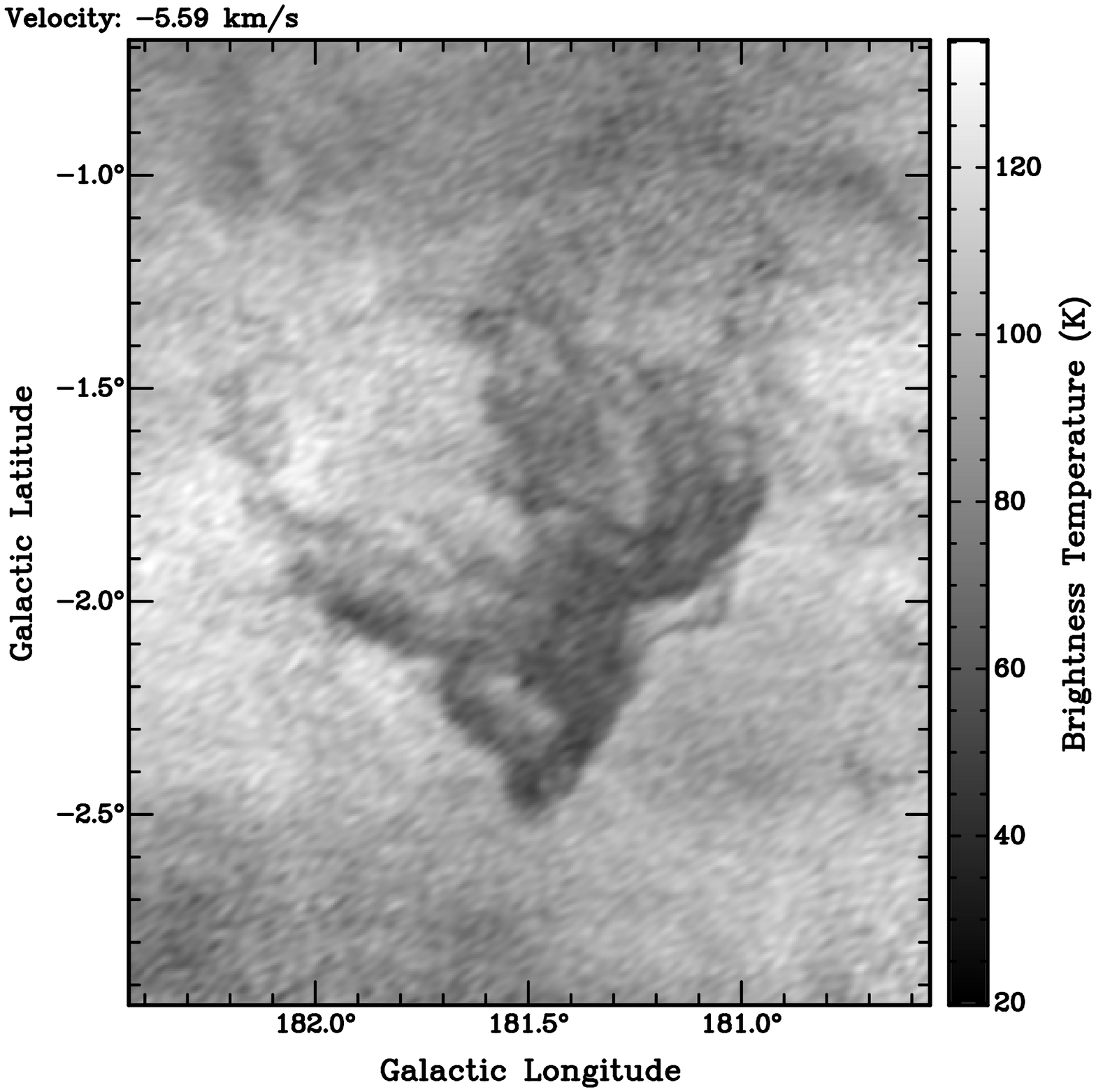}
}
\caption{
Striking anticenter 
{\sc H~i} self-absorption (HISA) 
features in new
%recently-released 
CGPS-III data, with
typical line widths $\Delta v \sim 2-3$km/s.  Neither has been identified
previously, except 
tentatively 
in the DRAO 26m {\sc H~i} survey
\citep{higgs_2005}.  Comparison to new FCRAO CO data is planned (see Brunt and
Mottram articles, this volume).
}
\label{Fig:cgps_hisa_ac}
\end{figure}

HISA is the preferred method for mapping detailed CNM structure in absorption
(Fig.~\ref{Fig:cgps_hisa_ac}).  Although not ubiquitous, bright {\sc H~i}
emission is smooth and extensive enough to allow HISA shadows of CNM clouds
to be imaged over large areas \citep{rc_1972,knapp_1974_kh3,wendker_1983}.
Most bright {\sc H~i} is near the Galactic plane, but this is also where the
bulk of the CNM is located \citep{cox_2005}.  HISA backgrounds are typically
not as bright as HICA backgrounds, so the CNM sampled by HISA has a lower
maximum $T_s$ than HICA and is more focused on the coldest {\sc H~i} where
H$_2$ formation is taking place.  In fact, very narrow-line HISA in H$_2$
clouds can probe the cloud chemistry and evolutionary state
\citep{li_2003,goldsmith_2005,goldsmith_2007}.  Lastly, since the HISA line
emission background must overlap in velocity with the absorbing cloud, HISA
radiative transfer samples both the temperature and velocity fields along the
line of sight, aiding Galactic structure investigations.  The rest of this
review discusses HISA results primarily.

%%%%%%%%%%%%%%%%%%%%%%%%%%%%%%%%%%%%%%%%%%%%%%%%%%%%%%%%%%%%%%%%%%%%%%%%%%%%%%
%%%%%%%%%%%%%%%%%%%%%%%%%%%%%%%%%%%%%%%%%%%%%%%%%%%%%%%%%%%%%%%%%%%%%%%%%%%%%%%

\section{HISA in the CGPS Era}
\label{Sec:cgps}

\begin{figure}[htb]
\centerline{
%\plottwo{gibson_steven_f3a_gray.eps}{gibson_steven_f3b_gray.eps}
%\plottwo{gibson_steven_f3a_color.eps}{gibson_steven_f3b_color.eps}
\plottwo{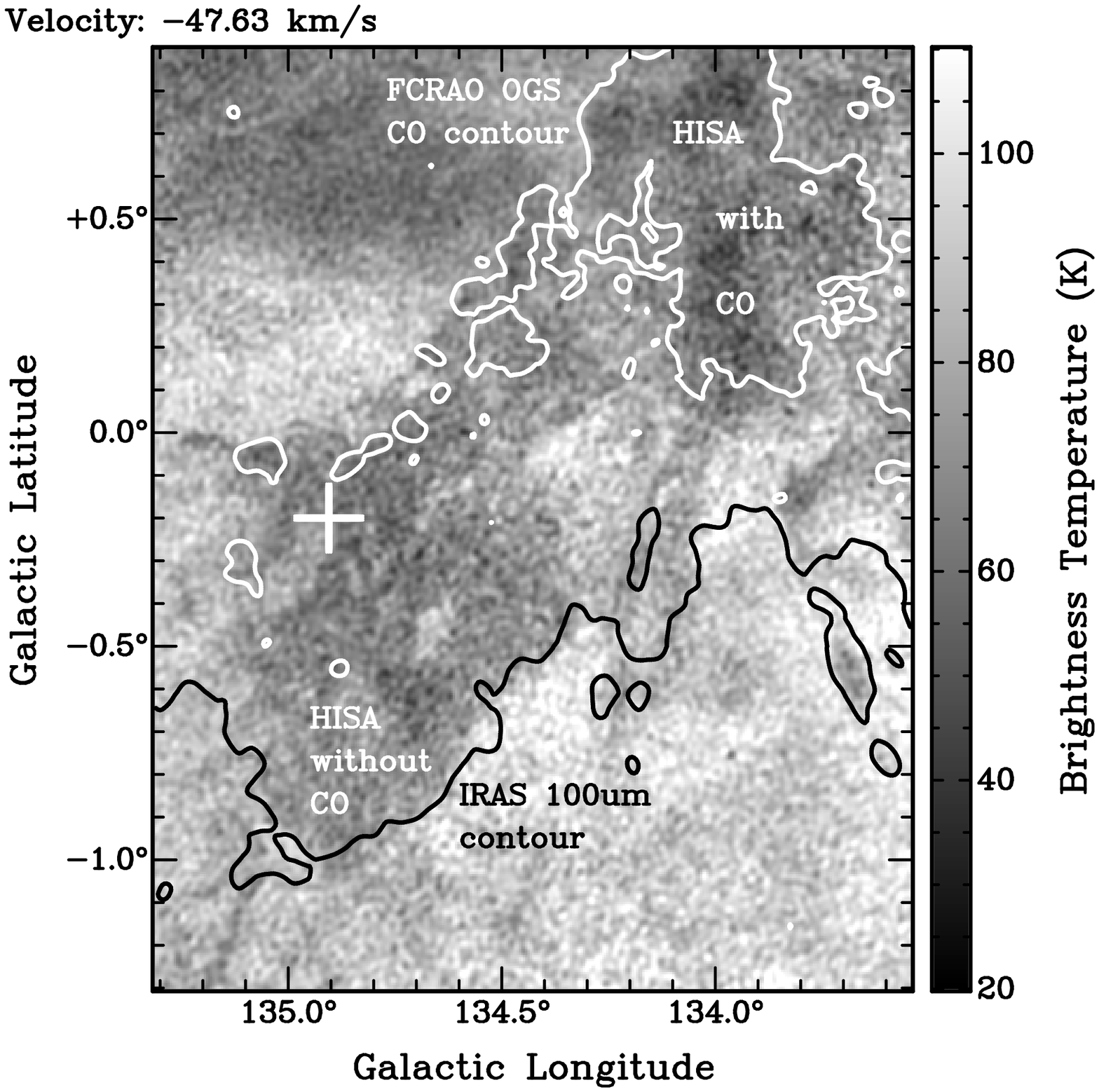}{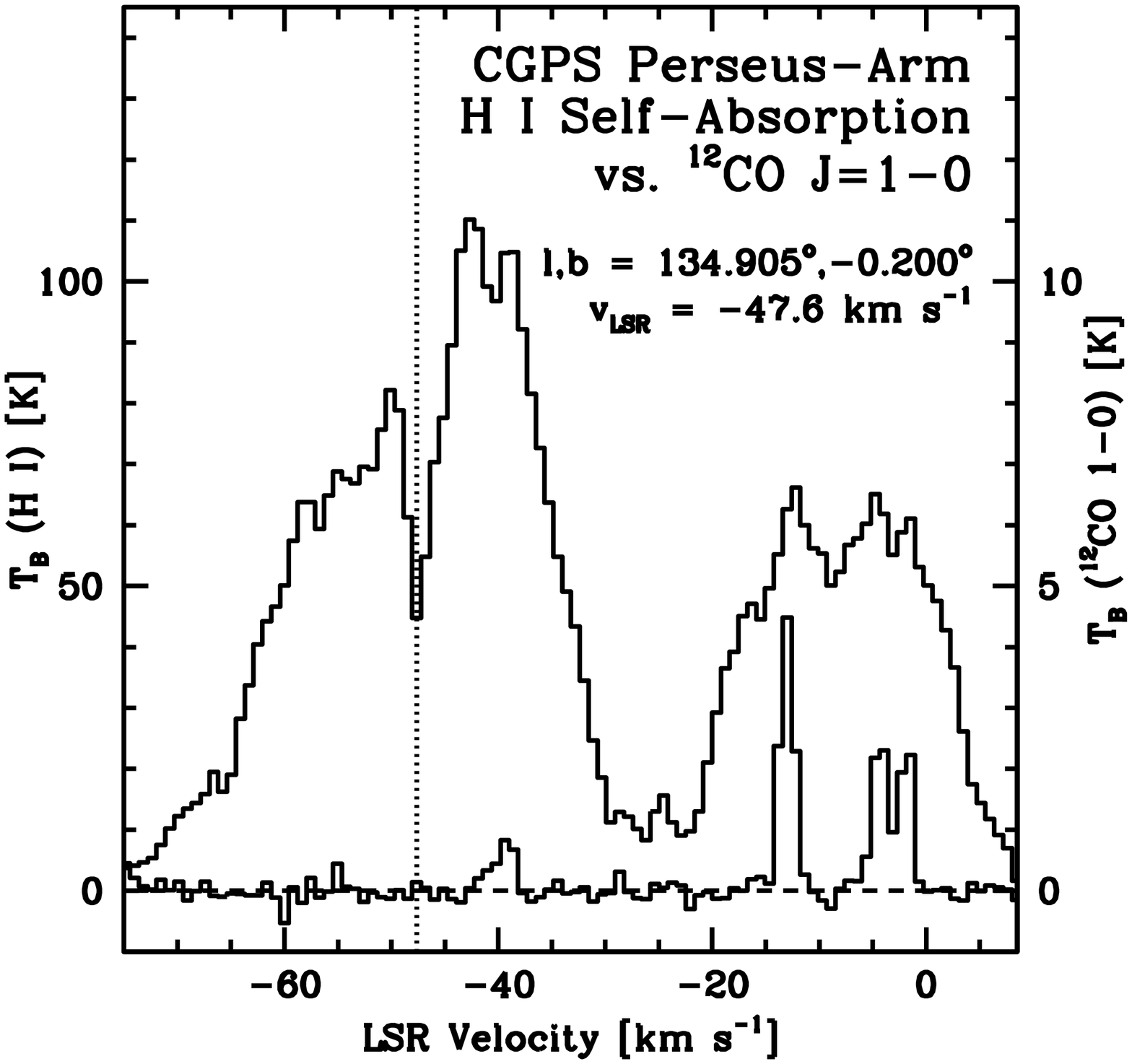}
}
\caption{
CGPS Perseus-arm HISA $\sim 1^\circ$ S of W4.  
{\it Left:\/} {\sc H~i} image with 
$^{12}$CO (0.6~K)
and 
FIR 100~$\mu$m (63~MJy/sr)
contours.  
The cross marks the position of the {\sc H~i} and CO spectra at right.  
HISA extends well beyond the detected CO but has matching dust emission,
suggesting H$_2$ untraced by CO.  This HISA-CO discrepancy is large enough to
be visible at $\sim 10'$ resolution \citep{hsf_1983}.
}
\label{Fig:cgps_hisa_w4}
\end{figure}

Before the CGPS, HISA was mostly used to examine {\sc H~i} content in molecular
clouds, either in single-beam observations toward lists of objects (e.g.,
\citealt{knapp_1974_survey,mccutcheon_1978}) or in synthesis imaging of small
areas (e.g., \citealt{landecker_1980,vanderwerf_1989}).  The common presumption
was
that HISA gas is too cold to exist without some form of molecular cooling and
shielding from the interstellar radiation field.  So time-consuming HISA
searches outside known dark/CO clouds were not pursued, except in
position-velocity strip surveys with Arecibo \citep{bb_1979,bl_1984} and in
lower-resolution maps of chance discoveries (e.g., 
\citealt{rc_1972,hsf_1983}).  
Many studies used the Maryland-Green Bank 91m {\sc H~i} survey \citep{ww_1982},
whose 
angular resolution was sufficient to study basic aspects of large HISA features
(e.g., Fig.~\ref{Fig:cgps_hisa_w4}).
The need for a synthesis imaging survey to detect and study HISA at smaller
scales was not widely appreciated.  However, such an undertaking was also
impractical before the computing power and automated techniques of the 1990s
\citep{higgs_1999}.

The CGPS was not designed as a HISA search engine, but its high angular
resolution and unbiased
coverage of a large area
proved ideal for this purpose.  New, intricate HISA clouds were 
found,
including many without corresponding CO emission (Fig.~\ref{Fig:cgps_hisa_w4};
see also \citealt{gibson_2000,knee_2001,kerton_2005}).  Similar discoveries
in the SGPS and VGPS came rapidly
\citep{mcclure_2001,kavars_2003,gibson_2004,mcclure_2006}.  All
were greatly aided
by the inclusion of single-dish data with the synthesis observations for full
$uv$ coverage.  In the CGPS,
comparison data in the $^{12}$CO $J=1-0$ line \citep{ogs} and far-infrared {\sl
  IRAS\/} dust emission \citep{iga,miga} were also a significant advantage.
The now-common view that large, blind, high-resolution, multiwavelength surveys
are worthwhile is due in no small part to the pioneering CGPS effort.

%%%%%%%%%%%%%%%%%%%%%%%%%%%%%%%%%%%%%%%%%%%%%%%%%%%%%%%%%%%%%%%%%%%%%%%%%%%%%%
%%%%%%%%%%%%%%%%%%%%%%%%%%%%%%%%%%%%%%%%%%%%%%%%%%%%%%%%%%%%%%%%%%%%%%%%%%%%%%%

%%%%%%%%%%%%%%%%%%%%%%%%%%%%%%%%%%%%%%%%%%%%%%%%%%%%%%%%%%%%%%%%%%%%%%%%%%%%%%
\begin{figure}[htb]
%\plotone{gibson_steven_f4_gray.ps}
%\plotone{gibson_steven_f4_color.ps}
\plotone{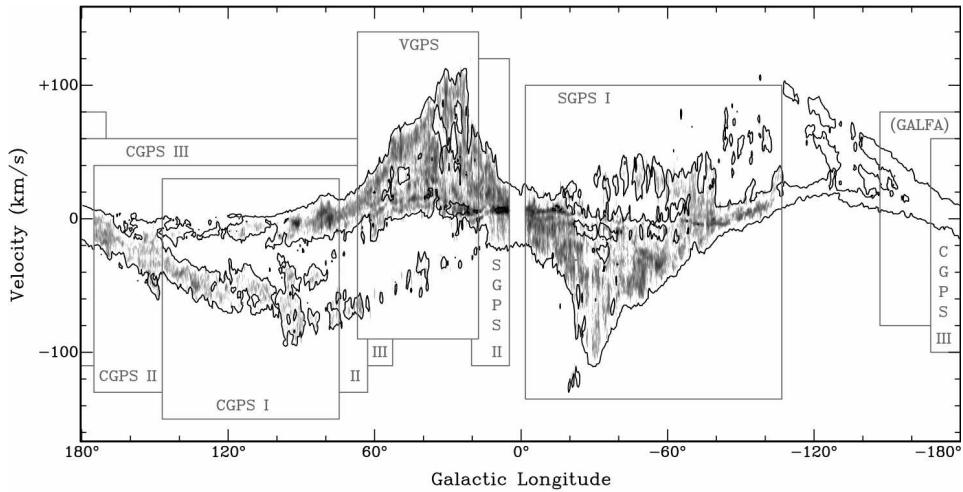}
\caption{
IGPS HISA longitude-velocity distribution over most of the Galactic disk,
integrated in latitude as $\int (T_{ON} - T_{OFF}) \, db$ (strong absorption
is dark).
%Various 
Survey areas are marked.  CGPS-III data are not yet included.  ``GALFA'' =
Arecibo anticenter coverage.
Contours show {\sc H~i} emission with $T_B = 70$~K, the minimum
for HISA to be detected reliably by the search algorithm.
Inner-Galaxy gas has $v_{LSR} \ga 0$~km/s for $0^\circ < \ell <
180^\circ$ and $v_{LSR} \la 0$~km/s for $180^\circ < \ell < 360^\circ$, while
gas outside the Sun's orbit occupies the other two $(\ell,v)$ quadrants.
}
\label{Fig:hisa_hie_lv}
\end{figure}
%%%%%%%%%%%%%%%%%%%%%%%%%%%%%%%%%%%%%%%%%%%%%%%%%%%%%%%%%%%%%%%%%%%%%%%%%%%%%%%

\section{HISA Galactic Distribution}
\label{Sec:dist}

HISA is very common at arcminute resolution \citep{gibson_2000,dickey_2003}.
The rich structure and sheer number of HISA clouds in the IGPS are too much to
analyze by hand, so automated methods of feature identification and
extraction were developed \citep{gibson_2005_meth,kavars_2005}.  Reliable
HISA identifications are distinguished from gaps in the general {\sc H~i}
emission by having narrow lines, fine angular structure, and smooth, bright
backgrounds.  These criteria select real features that appear as coherent
entities in ``movies'' of successive velocity channel maps in $(\ell,b,v)$
image cubes, often with significant kinematic structure.  One can also require
molecular/dust tracers for verification (e.g., \citealt{knapp_1974_survey}),
but these limit detections to the target clouds, missing the larger HISA
population.  Of course, the other criteria miss HISA with broader lines,
weaker backgrounds, etc., but such features are hard to distinguish
algorithmically from emission gaps.  They can be identified by eye
\citep{knee_2001} but cannot be included in any population study using
uniform criteria.

HISA surveys have been published for the Phase-I CGPS \citep{gibson_2005_res}
and SGPS \citep{kavars_2005}, and partially for the VGPS and CGPS-II
\citep{gibson_2004,gibson_2007}.  Fig.~\ref{Fig:hisa_hie_lv} shows the
longitude-velocity distribution from all these surveys, with the SGPS
reprocessed using the CGPS algorithm.  The HISA $(\ell,v)$ distribution looks
quite similar to its HICA equivalent (\citealt{strasser_2007}, Fig.~7).  Weak
self-absorption is found essentially everywhere that emission backgrounds are
bright enough, while stronger HISA is clumped into complexes along spiral
arms, tangent points, etc.  Abundant outer-Galaxy HISA and HICA are present
in the Perseus and Outer arms for $0^\circ < \ell < 180^\circ$, and in the
Sagittarius and Perseus arms for $180^\circ < \ell < 360^\circ$ (see
Fig.~\ref{Fig:hisa_co_lv} for arm labels).

The spatial distribution of IGPS HISA clouds is a function of the conditions
required to produce such cold {\sc H~i} and the radiative transfer geometry
needed for it to self-absorb.  In the outer Galaxy, where pure circular
rotation allows only one distance for a given velocity, the presence of HISA
requires other gas motions to provide the emission background.  Turbulent
eddies can make backgrounds for the scattered weak HISA, but spiral density
waves are needed for the more organized strong HISA \citep{gibson_2005_res}.
Cold {\sc H~i} may arise naturally downstream of spiral arm shocks
\citep{minter_2001,bergin_2004}, and colliding turbulent flows could make cold
clouds on smaller scales \citep{vazquez_2007}, so both mechanisms that reveal
cold {\sc H~i} as HISA may also be responsible for its creation.

In the inner Galaxy, individual spiral arms are difficult to discern
\citep{kavars_2005,gibson_2007}, either because cold interarm {\sc H~i} is
common, or because the arms themselves are less well separated in $(\ell,v)$.
Circular rotation allows two distances per velocity here, so near-side cold
{\sc H~i} automatically has a far-side emission background, whether in arms
or not.  This is probably why the inner-Galaxy HISA is more prominent and
widespread.  It is also the rationale behind using the detection of HISA in
CO clouds to resolve near/far kinematic distance ambiguities
\citep{jackson_2002,busfield_2006,anderson_2009,roman_2009}.  Such analyses
presume no far-side HISA occurs, or it is ``filled in'' by near-side
emission.  The former is unlikely given outer-Galaxy HISA (see also
Fig.~\ref{Fig:gmodel}), while the latter requires a conspiracy of
matching foreground NHIE features.  Near-side HISA may be so abundant that
these caveats are minor, but they should be assessed carefully.  The
additional implicit assumption that HISA only arises within CO clouds is also
problematic; selecting only HISA matching the shapes of CO clouds may help
\citep{anderson_2009}, but there is no guarantee of identical HISA+CO
morphology (e.g., Fig.~\ref{Fig:cgps_hisa_w4}).

\section{HISA Relation to Molecular Gas}
\label{Sec:co}

\begin{figure}[htb]
%\plotone{gibson_steven_f5_gray.ps}
%\plotone{gibson_steven_f5_color.ps}
\plotone{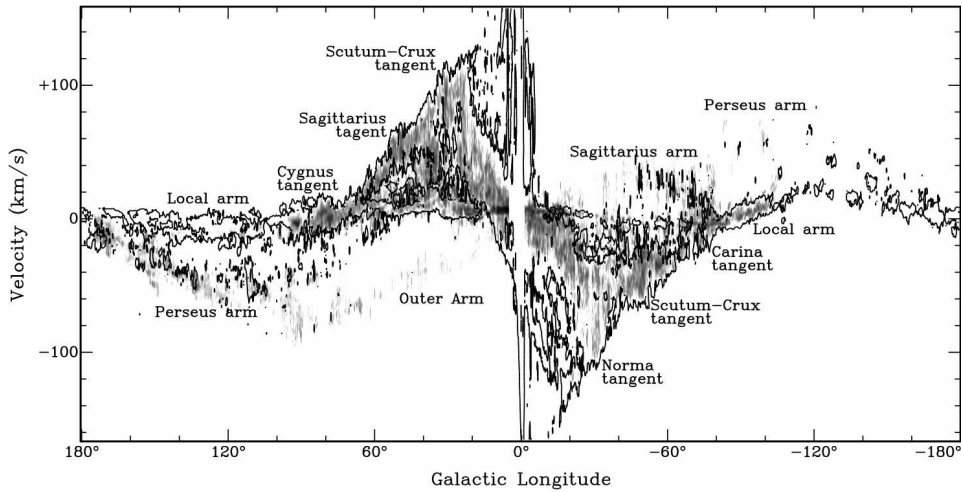}
\caption{
IGPS HISA $(\ell,v)$ distribution as in Fig.~\ref{Fig:hisa_hie_lv}, but with
\citet{cfa} CfA $^{12}$CO $J=1-0$ contours for latitude-integrated emission
$\int T_B \, db = 0.4$~K~deg, near the survey sensitivity limit.
Relevant spiral arms and tangent points are marked.
}
\label{Fig:hisa_co_lv}
\end{figure}

The IGPS HISA has a varying 
degree of correspondence with
CO emission
(Fig.~\ref{Fig:hisa_co_lv}).  Close inspection of HISA in the CGPS and VGPS
shows that {\it most inner-Galaxy HISA has matching CO, but most outer-Galaxy
  HISA does not\/}.
In the SGPS, which is dominated by inner-Galaxy HISA, only $\sim 60\%$ of
identified HISA clouds contain CO \citep{kavars_2005}.  CO without HISA is
readily explained as simply lacking the requisite bright {\sc H~i} emission
background.  The reverse case of HISA without CO is more interesting, as
standard CNM equilibrium models cannot explain very cold {\sc H~i} without
molecular gas \citep{wolfire_2003}.  Either the CNM models don't always apply,
or some HISA clouds have H$_2$ untraced by CO.

HISA without H$_2$ might arise if a shock removed the dust grains responsible
for photoelectric heating \citep{ht_2003}.  The next-most important heat
source, starlight photoionization of {\sc C~i}, would yield a much lower gas
temperature \citep{spitzer_1978,kh_1988,wolfire_1995}.  Oddly, such clouds
might be stable as very cold {\sc H~i}, since the same kind of dust dominates
both photoelectric heating and H$_2$ formation: very small grains and/or
aromatic hydrocarbons \citep{bakes_1994,habart_2004}.  Grains of this sort
may be destroyed in strong shocks \citep{ohalloran_2006,micelotta_2010}.
However, many shocks would be needed to account for all the weak HISA without
CO.
At the same time, the shocks could not be too large-scale, since many
stronger HISA features have partial CO correspondences
(Fig.~\ref{Fig:cgps_hisa_w4}).  So some CO-free HISA might be explained in
this way, but certainly not all.

%%%%%%%%%%%%%%%%%%%%%%%%%%%%%%%%%%%%%%%%%%%%%%%%%%%%%%%%%%%%%%%%%%%%%%%%%%%%%%
%%%%%%%%%%%%%%%%%%%%%%%%%%%%%%%%%%%%%%%%%%%%%%%%%%%%%%%%%%%%%%%%%%%%%%%%%%%%%%%

H$_2$ without CO emission can occur deep in molecular cores and in H$_2$ cloud
outer envelopes.
In cold cores, CO freezes onto grain mantles \citep{bergin_2007}.  But this
requires densities $\ga 10^5$~cm$^{-3}$, much higher than normal HISA gas
properties \citep{goldsmith_2005,klaassen_2005}, and CO would still be
visible in a lower-density region surrounding the core.
In molecular cloud envelopes, and diffuse molecular clouds generally, UV
absorption studies show that H$_2$ is shielded from dissociating UV photons
but CO is not for column densities of $N_H \sim 3-5 \times 10^{20}$~cm$^{-2}$
\citep{snow_2006,sheffer_2008}.
In addition, both the FCRAO Outer Galaxy and CfA Composite $^{12}$CO $J=1-0$
emission surveys \citep{ogs,cfa} have sensitivity cutoffs near $\sim
2$~K~km/s, or $N_H \sim 7 \times 10^{20}$~cm$^{-2}$ using the CfA conversion
factor.  Thus, current CO emission studies are blind to H$_2$-dominated gas
for $N_H \sim 3-7 \times 10^{20}$~cm$^{-2}$, which
may include a significant fraction of the total cloud mass
\citep{wolfire_2010}.  H$_2$ in this regime has been inferred
from gas property constraints in HISA clouds
\citep{hsf_1983,klaassen_2005,hosokawa_2007}, 
from ``infrared excess'' clouds
with more dust thermal radiation than their {\sc H~i} would imply
(\citealt{reach_1994,douglas_2007}), 
and 
from ``dark gas'' clouds with a similar excess of proton-scattered
$\gamma$-rays \citep{grenier_2005,abdo_2010}.  The widespread detection of
CO-free diffuse H$_2$ makes it a likely host for HISA.

Since many HISA clouds are near spiral density waves in longitude-velocity
space, the HISA-CO relationship may be {\it evolutionary\/} rather than
static.  In the grand-design view of star formation, diffuse atomic gas
entering a spiral arm is compressed in a spiral shock \citep{roberts_1969}.
Its sudden high density leads to rapid cooling \citep{spitzer_1978}, H$_2$
condensation \citep{koyama_2000,bergin_2004}, and star formation
\citep{roberts_1972,heyer_terebey_1998}.  This scenario has been proposed for
HISA in the CGPS \citep{gibson_2002,gibson_2005_res}, VGPS
\citep{minter_2001,gibson_2007}, and SGPS \citep{sato_1992,kavars_2005}
regions.  In outer-Galaxy arms, the HISA traces cold {\sc H~i} on the near
side of the arm and immediately downstream of the shock, where it is backlit
by warmer {\sc H~i} emerging on the far side of the arm
\citep{gibson_2005_res}.  The HISA and CO appear poorly mixed in all the
outer-Galaxy arms in Fig.~\ref{Fig:hisa_co_lv}, because (1)~the HISA may form
faster than the CO, or without sufficient UV shielding for CO, and (2)~much
of the CO may lie deeper within the spiral arm, with less {\sc H~i} emission
behind it for illumination.  Inner-Galaxy sight lines show much more CO with
HISA, since all near-side clouds have far-side backgrounds (see
\S~\ref{Sec:dist}).

Excitingly, \citet{braun_2009} have recently identified ``self-opaque'' cold
{\sc H~i} emission features in exquisitely sharp WSRT images of M31, which is
too inclined for cold {\sc H~i} to appear as HISA in the same fashion as in
the Milky Way.  Not only do many of the opaque {\sc H~i} clouds lie along the
edges of spiral arms near spiral shocks, but they also exhibit the same
incomplete correspondence with CO as IGPS HISA.

\section{Cold {\sc H~i} in Numerical Models}
\label{Sec:models}

%%%%%%%%%%%%%%%%%%%%%%%%%%%%%%%%%%%%%%%%%%%%%%%%%%%%%%%%%%%%%%%%%%%%%%%%%%%%%%
\begin{figure}[!b]
\centerline{
\includegraphics[width=1.65in]{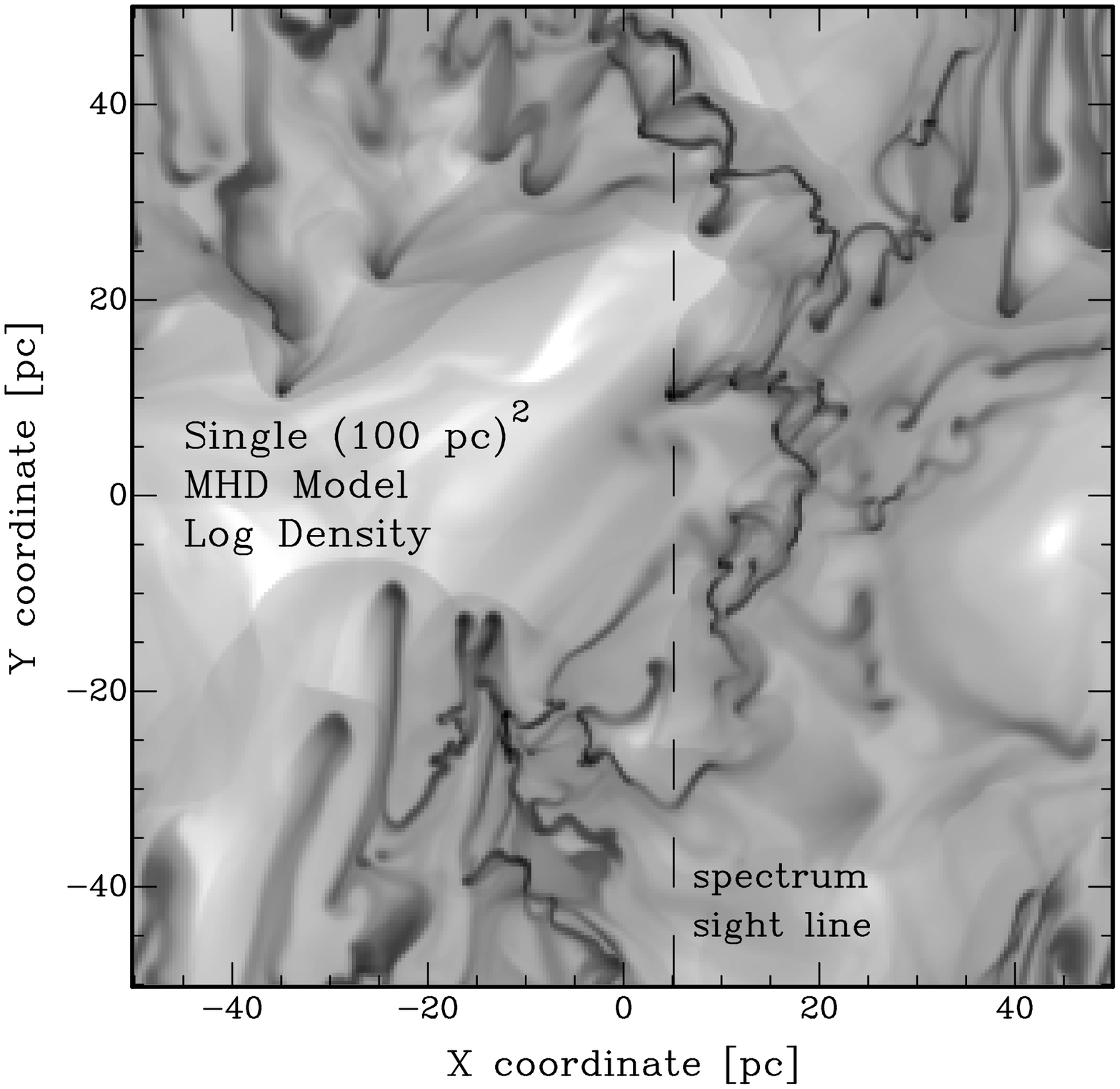}
\includegraphics[width=1.75in]{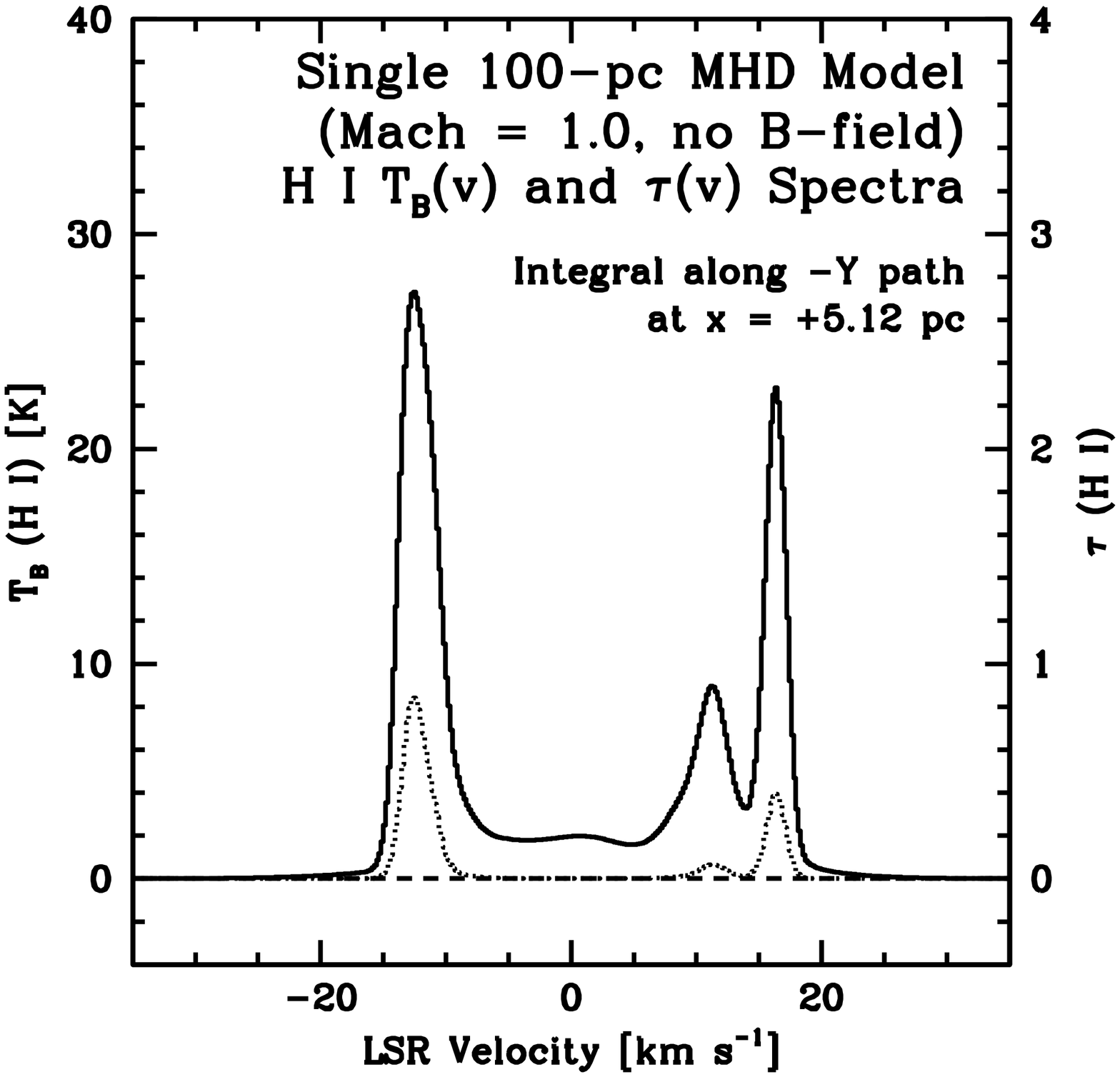}
\includegraphics[width=1.75in]{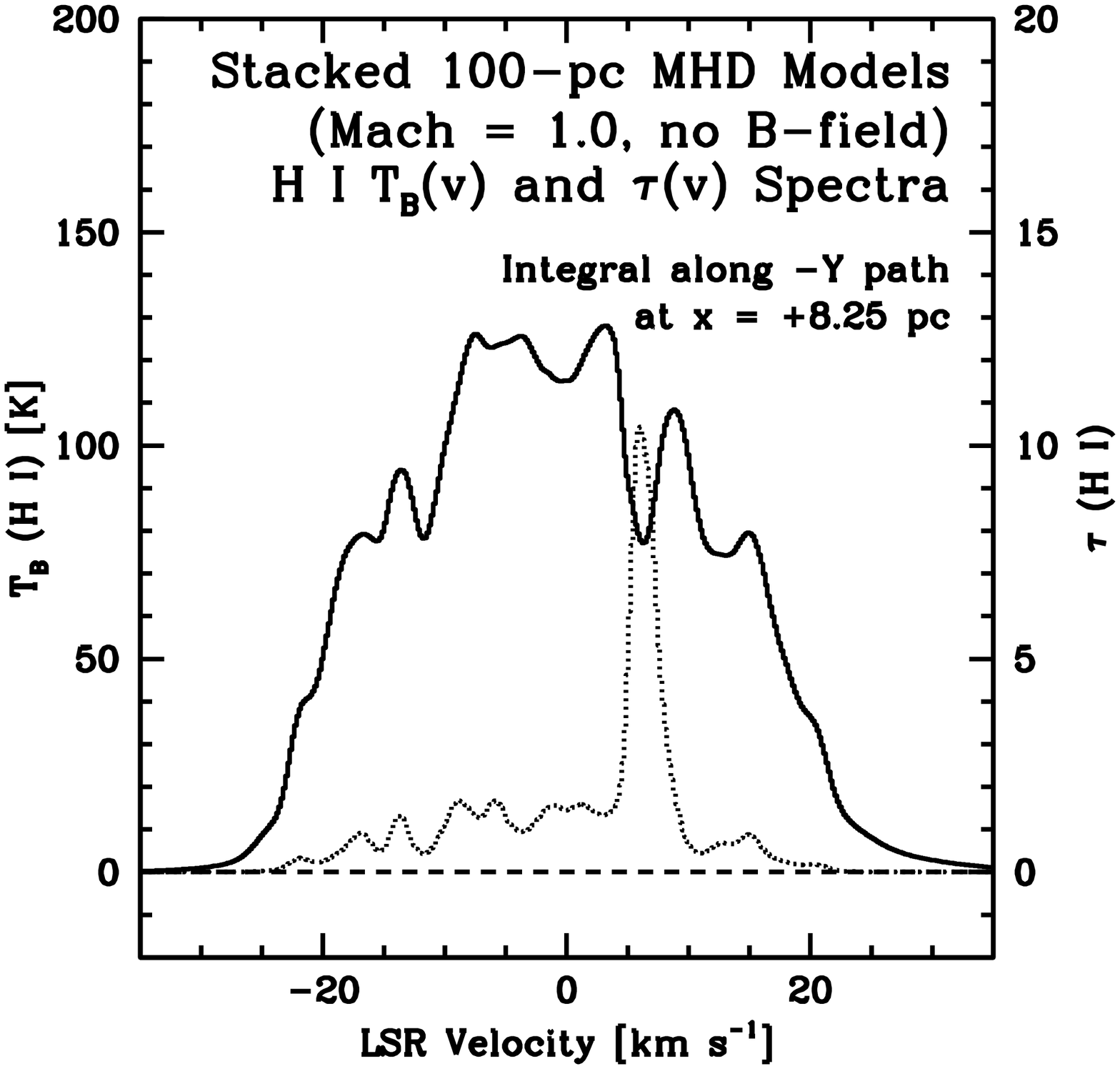}
}
\caption{
{\it Left:\/} $100 \times 100$~pc$^2$ 2-D MHD model log(density) field
(0.05~pc cells).
{\it Center:\/} Synthetic observed spectrum through the model.  NHIE features
appear as matching peaks in $T_B$ and $\tau$.
{\it Right:\/} Synthetic spectrum from 40 concatenated models.  This
simulates a much larger ISM column, providing adequate backgrounds for HISA,
in which $T_B$ dips coincide with $\tau$ peaks.
}
\label{Fig:hmodel}
\end{figure}
%%%%%%%%%%%%%%%%%%%%%%%%%%%%%%%%%%%%%%%%%%%%%%%%%%%%%%%%%%%%%%%%%%%%%%%%%%%%%%%

The computing power that enabled the IGPS also allowed significant advances
in ISM numerical models at scales of interstellar clouds
\citep{vazquez_2006,vazquez_2007,hennebelle_2007_prop,hennebelle_2008} and
spiral arms \citep{dobbs_2006,dobbs_2008,kim_2006,kim_2008}.  While these new
models are very interesting, only a few have been put in a form that can
readily be compared to data from a real radio telescope (e.g.,
\citealt{hennebelle_2007_obs,douglas_2010}).  This is accomplished through
``synthetic observations'' of numerical models inside the computer, applying
radiative tranfer to $(x,y,z)$ grids of density, temperature, and velocity to
produce $(\ell,b,v)$ brightness temperature grids of, e.g., {\sc H~i} 21cm line
emission.  It is natural to wonder how well CNM tracers like NHIE
and HISA can be discerned in such simulations, and whether comparisons to
real observations can constrain the model physics.

%%%%%%%%%%%%%%%%%%%%%%%%%%%%%%%%%%%%%%%%%%%%%%%%%%%%%%%%%%%%%%%%%%%%%%%%%%%%%%
\begin{figure}[!ht]
%\plotone{gibson_steven_f7a_gray.eps}
%\plottwo{gibson_steven_f7b_gray.eps}{gibson_steven_f7c_gray.eps}
%\plotone{gibson_steven_f7a_color.eps}
%\plottwo{gibson_steven_f7b_color.eps}{gibson_steven_f7c_color.eps}
\plotone{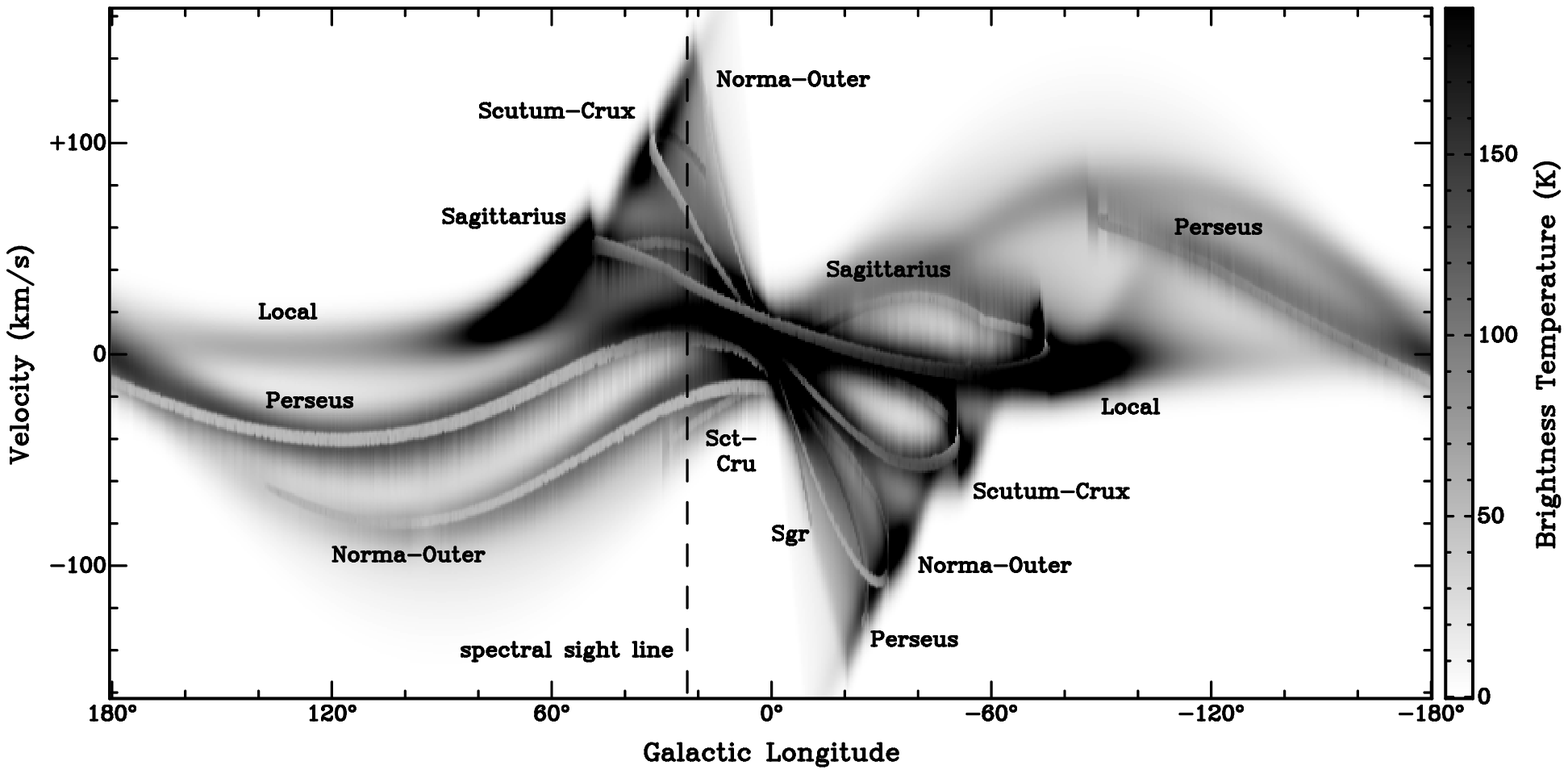}
\plottwo{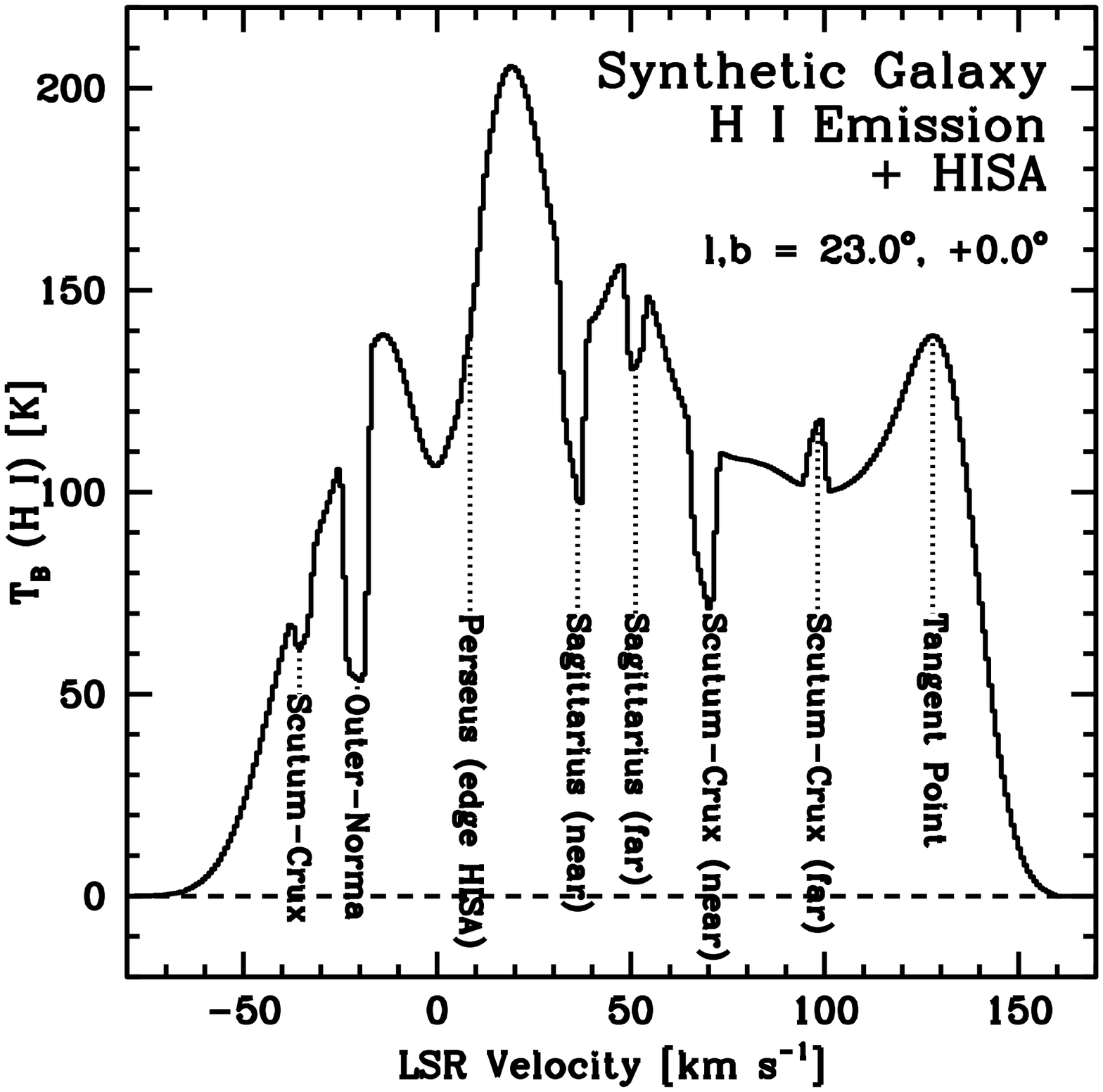}{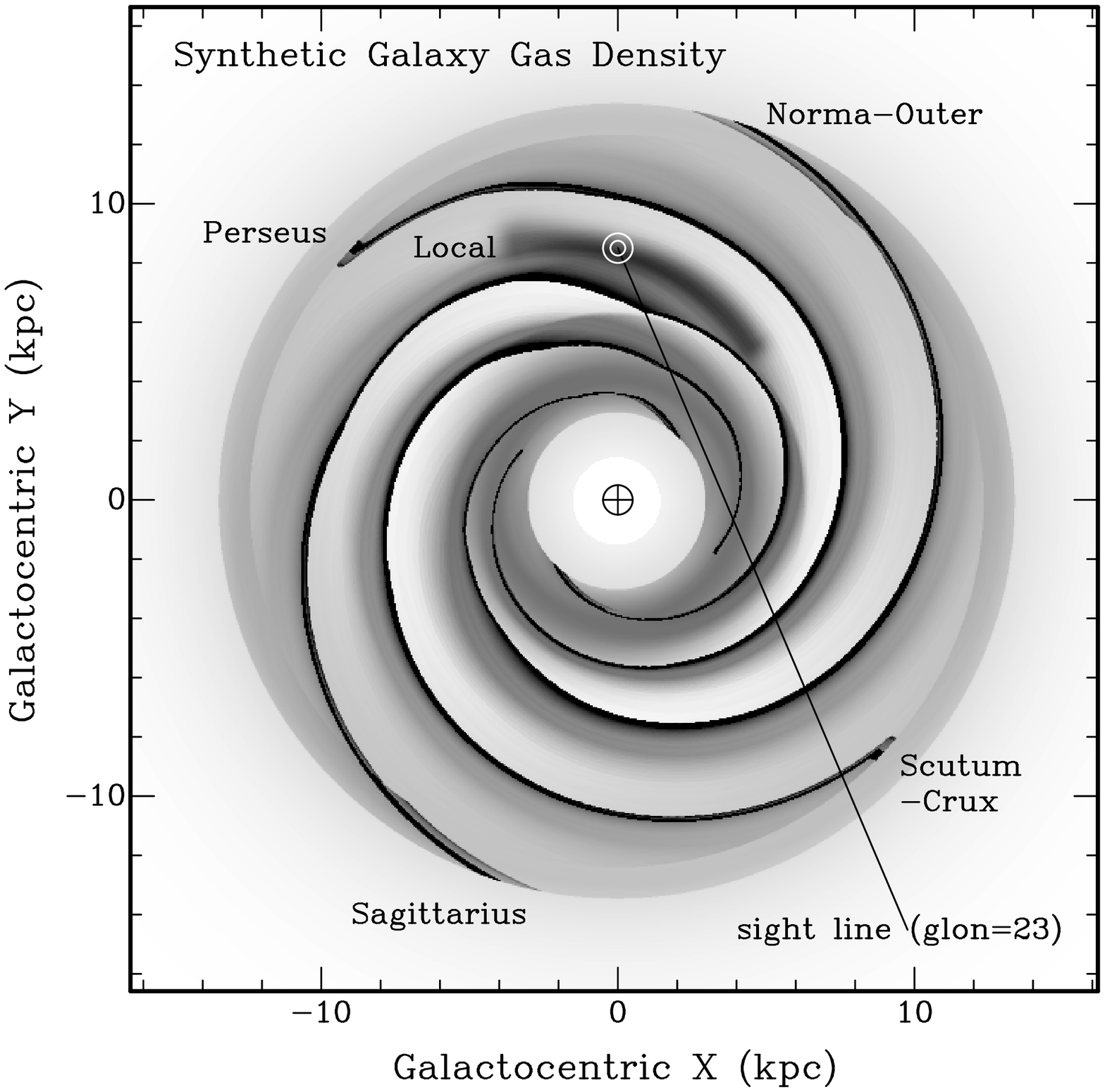}
\caption{
{\it Top:\/} Synthetic $(\ell,v)$ observations of a simple 2-D Galactic {\sc
  H~i} model with cold {\sc H~i} downstream of spiral shocks, where it is
visible as HISA 
%\citep{gibson_gmodel}.  
(Bell \& Gibson, in prep).
The intensity scale is negative, so
absorption appears light.
{\it Lower Left:\/} Spectrum at $\ell = 23^\circ$, showing HISA in
outer-Galaxy arms and both near- and far-side inner-Galaxy arms; one
Scutum-Crux arm crossing appears in NHIE.
{\it Lower Right:\/} Plan-view of the model density distribution (dark is
more dense).
}
\label{Fig:gmodel}
\end{figure}
%%%%%%%%%%%%%%%%%%%%%%%%%%%%%%%%%%%%%%%%%%%%%%%%%%%%%%%%%%%%%%%%%%%%%%%%%%%%%%%

%\citet{gibson_hmodel} 
Gibson et al.\/ (in prep)
have made synthetic observations of $100 \times 100$~pc$^2$
magneto-hydrodynamic (MHD) models of gas with a wide range of densities and
temperatures from \citet{gazol_2005,gazol_2009}, to see (a) under what
conditions cold {\sc H~i} can be identified as HISA or NHIE, and (b) how well
the appearances of these features under different model conditions match real
HISA and NHIE clouds.  As Fig.~\ref{Fig:hmodel} shows, NHIE features are
common, with amplitudes and line widths similar to CGPS NHIE
(Fig.~\ref{Fig:cgps_nhie}), but there is not enough {\sc H~i} emission
background for HISA.  This can be crudely addressed by observing many models
end-to-end to simulate a larger region (Fig.~\ref{Fig:hmodel}), although this
requires a considerable path length if the mean model density (1~cm$^{-3}$)
is not increased.  Curiously, even with many models added together, the
classical WNM gas with $T > 1000$~K does not produce enough background
emission by itself for the HISA to be seen.  The rest of the emission comes
from cooler gas, including thermally unstable gas between the equilibrium WNM
and CNM regimes.  Such unstable gas, predicted by prior simulations
\citep{gazol_2001}, has been found in great quantity in recent HICA studies
\citep{heiles_2001,ht_2003}.

An alternative approach is to model the whole Galaxy, including density,
temperature, and velocity variations in spiral arms.  Fig.~\ref{Fig:gmodel}
shows an example of this with cold {\sc H~i} in bands downstream of
spiral shocks (\citealt{gibson_2006_model_poster}; Bell \& Gibson, in prep).
The arm positions and shock parameters are not intended to be exact, as the
goal is merely to seek qualitative agreement with the HISA seen in the IGPS.
The arm pattern is adapted from \citet{tc_1993}, with a \citet{wolfire_2003}
global {\sc H~i} distribution and flat rotation curve modified by
\citet{roberts_1969,roberts_1972} spiral shocks.  Each 25~pc $(x,y)$ model
cell contains either pure WNM ($T=8000$~K, $n \sim 0.5$~cm$^{-3}$) or pure
CNM ($T=40$~K, $n \sim 100$~cm$^{-3}$), with the latter usually confined to
the major spiral arms (not the Local arm in this example).  This simple
model produces copious HISA in spiral arms all across the Galactic disk
(Fig.~\ref{Fig:gmodel}).
The WNM filling factor is fudged here (100\%), but emission backgrounds are
too weak to make HISA with the local value ($\sim 50\%$; \citealt{ht_2003}).
As with the MHD model above, this may indicate that real HISA backgrounds
arise from a mix of classical WNM gas and cooler, more opaque {\sc H~i}
emission.
For clarity, the sample model in Fig.~\ref{Fig:gmodel} has no turbulence, nor
any cold {\sc H~i} between arms.  Experiments with different models indicate
that the IGPS outer-Galaxy HISA is best explained by turbulent CNM in arms
rather than random clouds throughout the disk, while inner-Galaxy HISA
requires at least some interarm CNM, or arms that are less clearly separated
in radial velocity \citep{gibson_2007}.

\citeauthor{douglas_2010} (\citeyear{douglas_2010}; Douglas, this volume) use
a more sophisticated approach with 3-D $(\ell,b,v)$ synthetic observations of
the Galactic-scale MHD models of \citet{dobbs_2008} to simulate the CGPS {\sc
  H~i} data set.  They find considerable HISA in their own version of the
Perseus arm and are able to track throughout the radiative transfer to
determine exactly where the HISA arises and under what conditions.

\section{Magnetic Fields?}
\label{Sec:magnetic}

%%%%%%%%%%%%%%%%%%%%%%%%%%%%%%%%%%%%%%%%%%%%%%%%%%%%%%%%%%%%%%%%%%%%%%%%%%%%%%
\begin{figure}[!b]
\centerline{
\includegraphics[width=2.5in,angle=-90]{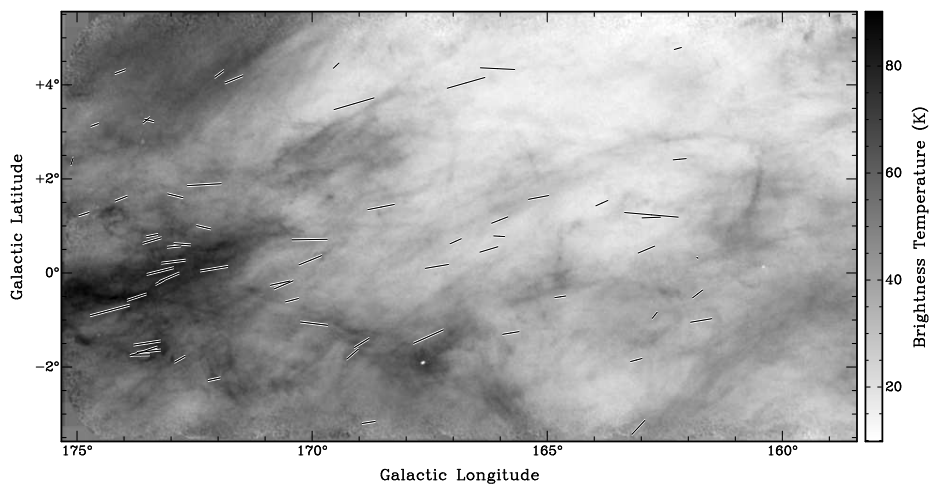}
}
\caption{
CGPS {\sc H~i} emission filaments in local gas ($v_{LSR} = +5$~km/s).
Lines show interstellar magnetic field direction from starlight polarization
data \citep{heiles_2000}.
A line 1$^\circ$ long represents 5\% optical polarization.
}
\label{Fig:mag}
\end{figure}
%%%%%%%%%%%%%%%%%%%%%%%%%%%%%%%%%%%%%%%%%%%%%%%%%%%%%%%%%%%%%%%%%%%%%%%%%%%%%%%

The influence of magnetic fields on ISM structure is unclear but may be
significant 
%\citep{cox_2005}.  
\citep{kh_1988}.  Fig.~\ref{Fig:mag} shows a possible alignment of {\sc H~i}
emission filaments and $\vec{B}$-field direction measured from starlight
polarization.  The filaments are near the detection limits for $3'$ smoothed
CGPS data ($\Delta T_B \sim 5$~K), but those that can be isolated in velocity
are cold ($\Delta v \sim 5$~km/s), with column densities similar to small
HISA features ($N_H \sim 5 \times 10^{19}$~cm$^{-2}$ for $\tau << 1$).  Many
such alignments are visible over the survey area.  Is the field following the
filaments, or vice-versa?  \citet{mcclure_2006} found a similar alignment in
stunning SGPS images of the \citet{rc_1972} HISA cloud.  Following their
analysis using the RMS scatter of position angles $\sigma_{PA}$
\citep{cf_1953}, the predicted CGPS mean sky-plane field strength is

\begin{equation}
	\langle B_\perp \rangle  \; = \;
		       63.5 \, \mu {\rm G} \,
		       \left ( \frac{10^\circ}{\sigma_{PA}} \right)
		       \sqrt{ 
			  \left( \frac{100 \, {\rm pc}}{d} \right)
			  \left( \frac{3'}{\Delta \theta} \right)
			  \left( \frac{\Delta T_B}{5 \, {\rm K}} \right)
			  \left( \frac{\Delta v_{turb}}{5 \, {\rm km/s}} \right)
			}
	\;\; .
\end{equation}

\noindent This is an order of magnitude more than expected for the diffuse
ISM, but it's consistent with the \citet{mcclure_2006} HISA and some other
cases (e.g., \citealt{andersson_2005}).  Could some CNM gas have strong
$B$-fields?  More sensitive Arecibo data show similar alignments for 
%much
fainter filaments ($\Delta T \ga 0.5$~K; J. Peek, in prep) that may probe
more ordinary $B$-field strengths of a few $\mu$~G.

\section{GALFA and Future Surveys}
\label{Sec:galfa}

\begin{figure}[tb]
\centerline{
\includegraphics[width=2.60in]{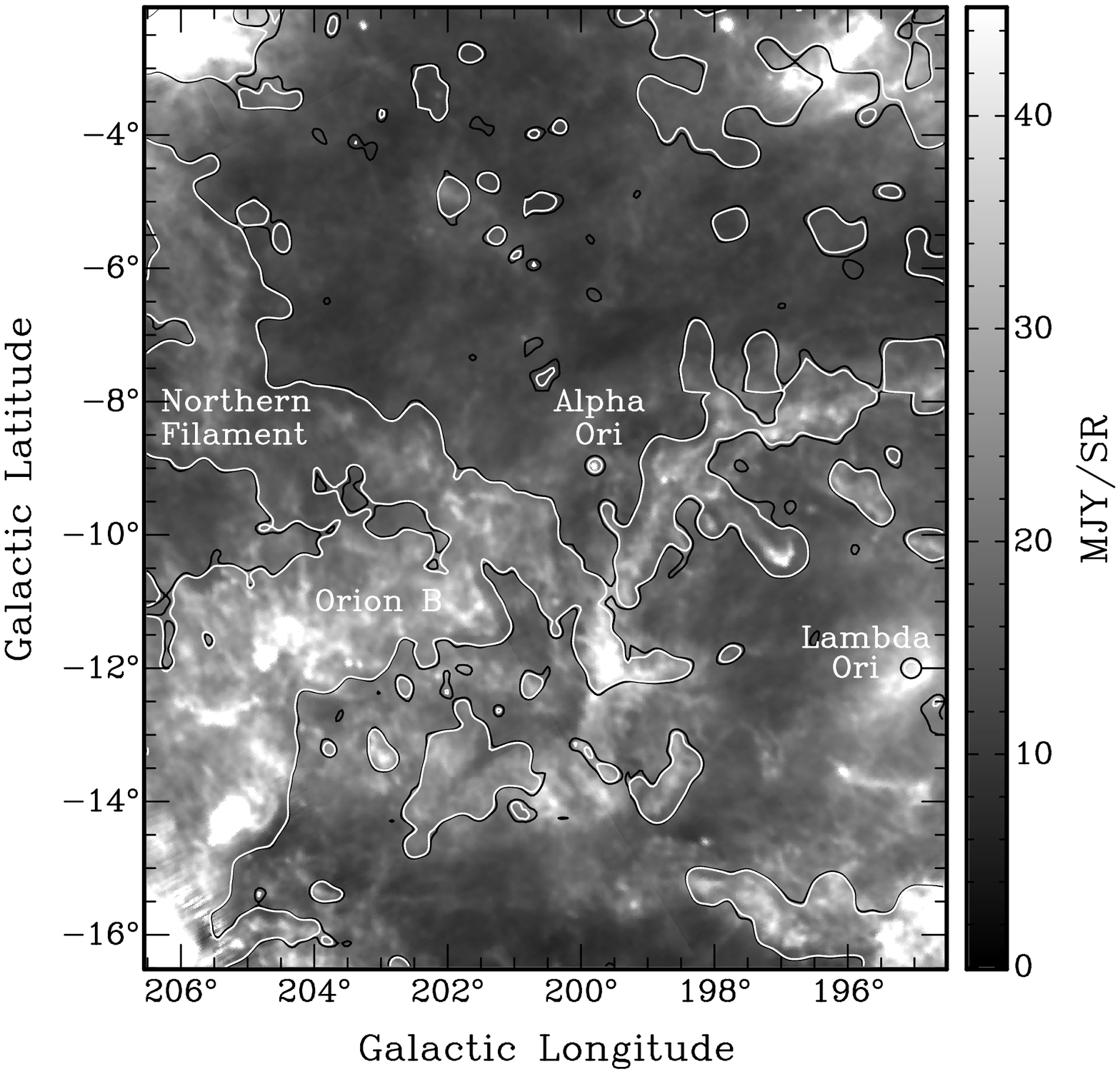}
\includegraphics[width=2.63in]{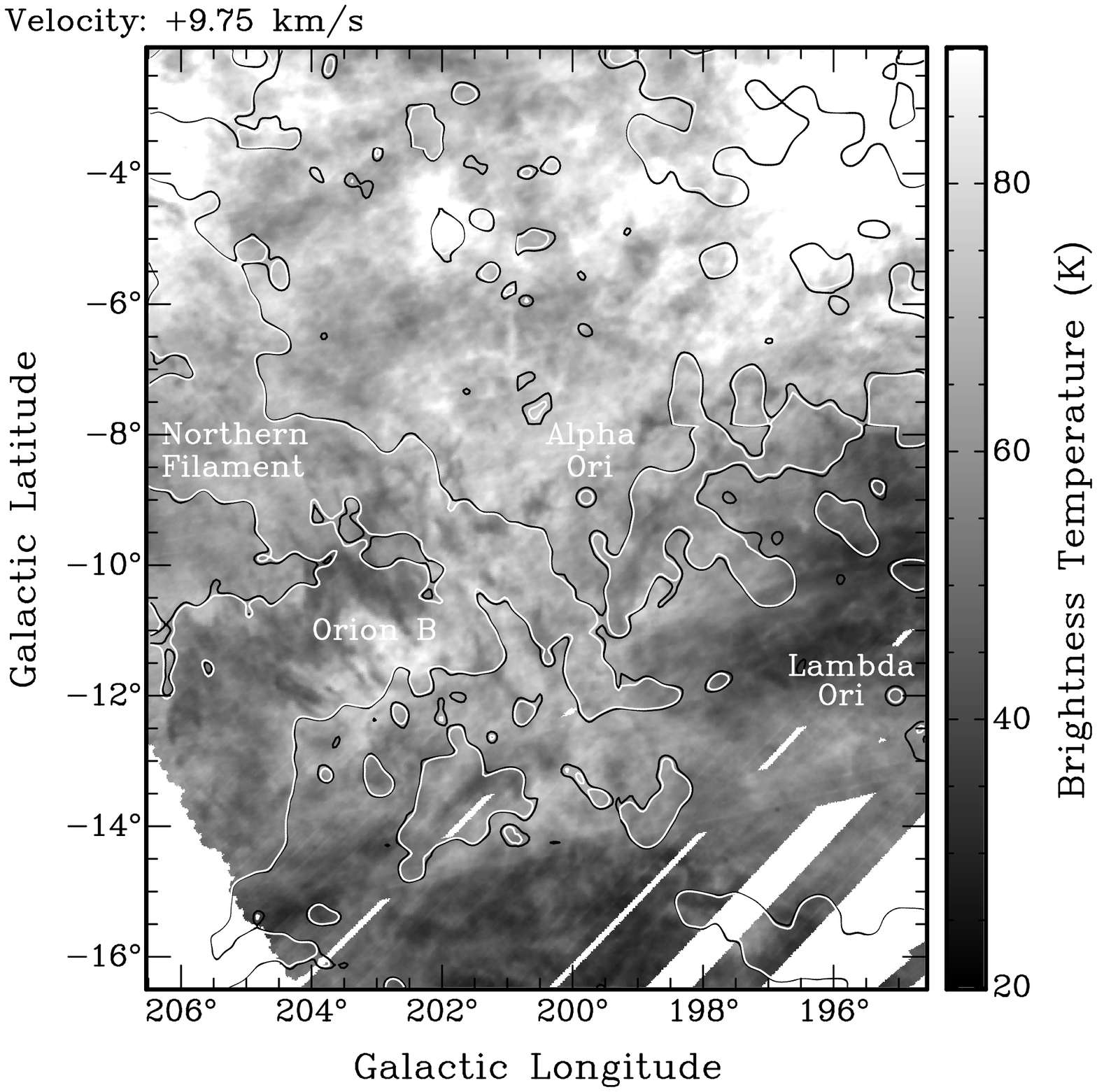}
}
\caption{
The northeast part of Orion.  {\it Left:\/} IRAS 100~$\mu$m dust emission;
{\it Right:\/} GALFA {\sc H~i} at $+10$~km/s (not fully mapped).  Contours
are CfA CO 1-0 line integral at 1~K~km/s.  Stars $\alpha$ and $\lambda$~Ori
are marked, as are the Ori~B and N. Filament molecular complexes; the $\sim
8^\circ$ diameter $\lambda$~Ori molecular ring can also be seen.
Richly-detailed HISA is visible in many CO clouds and outside a few (e.g.,
near $\alpha$~Ori).  HISA has been detected previously in this region
\citep{mccutcheon_1978,wannier_1983,sandqvist_1988,li_2003} but has never
been imaged as above.
}
\label{Fig:galfa_orion}
\end{figure}

The IGPS surveys have transformed our view of the CNM 
at arcminute scales, but they are hampered by 20th-century interferometer
sensitivities.  Another approach is to use the 
Arecibo 305m telescope to map cold
{\sc H~i} with a slightly larger beam ($3.4'$) but much better sensitivity.
The latter, along with installation of
the
ALFA 7-beam feed,
has allowed better velocity sampling and more rapid mapping in the Galactic
ALFA (GALFA) {\sc H~i} survey (see Peek, this volume).  GALFA targets the
whole Arecibo sky ($-1.3^\circ < \delta < 37.9^\circ$; 32\% of 4$\pi$ sr),
including the 1st quadrant and anticenter in the plane and a wide swath of
high-latitude gas.  GALFA's sensitivity, resolution, and sky coverage are
thus highly complementary to the IGPS.  A great many beautiful cold {\sc H~i}
features are visible in GALFA data (e.g., Fig.~\ref{Fig:galfa_orion}).

%%%%%%%%%%%%%%%%%%%%%%%%%%%%%%%%%%%%%%%%%%%%%%%%%%%%%%%%%%%%%%%%%%%%%%%%%%%%%%
%%%%%%%%%%%%%%%%%%%%%%%%%%%%%%%%%%%%%%%%%%%%%%%%%%%%%%%%%%%%%%%%%%%%%%%%%%%%%%%

Next in line is GASKAP, the Galactic spectral line survey with the Australian
SKA Pathfinder (Dickey, this volume).  By incorporating new array-feed
technology on an interferometer, ASKAP combines the field-of-view of small
dishes like DRAO with better resolution, speed, and bandwidth --- enough to
capture both {\sc H~i} and OH emission, so that the CNM and a hitherto
elusive tracer of the molecular medium can be surveyed together for the first
time.  GASKAP will image the whole Galactic plane within $|b| < 10^\circ$
and $\delta < +40^\circ$ at $10 - 20''$ resolution, yeilding an
unprecedentedly rich panorama of the dynamic ISM in our home galaxy.  Further
down the road, the SKA will enable sensitive {\sc H~i} imaging at
few-arcsecond scales, matching photographic sky surveys at last, and enabling
studies of nearby Galaxies at the same level of detail the IGPS pioneered for
the Milky Way.  Both ASKAP and the SKA will also be HICA machines, capturing
a huge grid of gas measurements that will revolutionize our view of the
Galaxy yet again.
The future looks very bright for studies of ``dark'' gas!

\acknowledgements 

I am grateful for the interaction and assistance of many more collaborators,
students, and observatory staff than can be listed here.
Karma visualization software
%\footnote{See also \url{http://www.atnf.csiro.au/karma}.}
\citep{karma}, 
the SuperMongo plotting package \citep{sm},
and the ADS abstract service were used extensively for this work.
Funding was provided by 
Western Kentucky University, the U.S. NSF, and Canadian NSERC.

%\bibliography{gibson_steven}
\bibliography{gibsonsj}

\end{document}